\documentstyle[12pt,aaspp4]{article}
\def\beq{\begin{equation}}
\def\eeq{\end{equation}}
\def\ref{\reference}
\def\simge{\mathrel{%
   \rlap{\raise 0.511ex \hbox{$>$}}{\lower 0.511ex \hbox{$\sim$}}}}
\def\simle{\mathrel{
   \rlap{\raise 0.511ex \hbox{$<$}}{\lower 0.511ex \hbox{$\sim$}}}}
\eqsecnum
\begin{document}
\title{Signatures of Energetic Protons in Hot Accretion Flows: Synchrotron
Cooling of Protons in Strongly Magnetized Pulsars}
\author{Jongmann Yang$^{1,2}$ and Insu Yi$^{1,2,3,4}$}
\affil{$^1$Center for High Energy Astrophysics and Isotope Studies, 
Research Institute for Basic Sciences}
\affil{$^2$Department of Physics, Ewha University, Seoul, Korea;
jyang@astro.ewha.ac.kr, yi@astro.ewha.ac.kr}
\affil{$^3$Korea Institute for Advanced Study, Seoul, Korea}
\affil{$^4$Institute for Advanced Study, Princeton, NJ 08540}

\begin{abstract}
The existence of hot, two-temperature accretion flows is essential to
the recent discussions of the low luminosity, hard X-ray emission from
accreting neutron stars and black holes in Galactic binaries and massive
black holes in low luminosity galactic nuclei. In these flows, protons are
essentially virialized and relativistic energies for non-thermal protons
are likely. Observational confirmation of the energetic protons' presence
could further support the two-temperature accretion flow models.
We point out that synchrotron emission from nonthermal relativistic 
protons could provide an observational signature in strongly magnetized 
neutron star systems. The self-absorbed synchrotron emission from an 
accreting neutron star with the magnetic moment $\sim 10^{30}~Gcm^3$ is 
expected to exhibit a spectrum $\nu I_{\nu}\sim \nu^2$ with the luminosity 
$\sim$ a few $\times 10^{33} (L_x/10^{36} erg/s)^{0.4}~erg/s$ at 
$\nu\sim 10^{15} Hz$ where $L_x$ is the X-ray luminosity from the 
neutron star surface. The detection of the expected synchrotron signature 
in optical and UV bands during the low luminosity state of the pulsar 
systems such as 4U 1626-67 and GX 1+4 could prove the existence of the 
hot, two-temperature accretion flows during their spin-down episodes.
The detected optical emission in 4U 1626-67 has a spectral shape and
luminosity level very close to our predictions.

\end{abstract}


\section{Introduction}

It has recently been suggested that the accretion flows around black holes
and neutron stars are very hot and two-temperatured when their luminosities 
are low and spectra are hard (e.g. Narayan \& Yi 1995, Rees et al. 1982 and
references therein). In such flows, often called advection-dominated
accretion flows (ADAFs), 
most of the viscously dissipated energy is used to heat ions 
and only a small fraction of the total energy is radiated by electron cooling 
processes. X-ray and gamma-ray emission properties in Galactic X-ray 
transients (e.g. Narayan et al. 1998b for a review) and galactic nuclei 
(e.g. Yi \& Boughn 1998 and references therein) including Sgr A$^*$ 
(Manmoto et al. 1997, Narayan et al. 1998a) have been successfully modeled 
by these hot, two-temperature flows. The apparent success of these models 
critically relies on the existence of the two-temperature plasma in which 
the ion temperature is essentially the virial temperature and hence much
higher than the electron temperature (Narayan \& Yi 1995). Unless there is 
an efficient electron-ion energy exchange mechanism, the Coulomb exchange 
alone typically leads to the two-temperature condition (Narayan et al. 1998b). 

Although the spectral signatures seen in various radiation components due to 
electrons at temperatures $\sim 10^9-10^{10}K$ are quite plausible, the 
electron spectral components alone cannot prove the uniqueness of the 
spectral fits in the X-ray systems studied so far. Any direct observational 
signatures due to energetic protons could be extremely useful to confirm the 
presence of the protons and hence the presence of the two-temperature 
accretion flows. However, direct proton radiation signatures are difficult 
to observe since the radiation efficiencies of proton-related radiation 
processes are usually very low (Mahadevan 1998). There have been some recent 
suggestions that energetic ions could provide observable signatures through 
various phenomena such as pion production (Mahadevan et al. 1997, Mahadevan 
1998) and nuclear spallation (Yi \& Narayan 1997). 
It is however unclear how the suggested
possibilities are proved as unambiguous signatures of the energetic protons
in the two-temperature plasma (Yi \& Narayan 1997).
It is therefore interesting to see whether there are any other signatures of 
the two-temperature accretion flows in which protons have relativistic 
energies. In this paper, we point out that there could be such a signature 
produced by the proton synchrotron emission in the strong magnetic fields 
around neutron stars. Since electrons are rapidly cooled near the neutron 
stars by soft photons from the stellar surface, the electron synchrotron 
signature is not expected (Yi \& Narayan 1995).

Within the accretion flows, since the proton gyroradius 
$\sim 3\gamma_2 B_8^{-1}~cm$ is much smaller than the length scale of the 
accretion flow, the protons are likely to be tightly bound within the 
accretion flows where $B_8=B/10^8~G$ is the magnetic field strength and 
$\gamma_2=\gamma/10^2$ is the Lorentz factor for relativistic protons. 
The characteristic synchrotron loss time scale is
$t_{sync}\sim 5\gamma_2^{-1}B_8^{-2}~s$. The typical accretion time
scale in the hot ADAFs $t_{acc}\sim 3\times 10^{-5} \alpha^{-1}m r^{1/2}~s$
where $m=M/M_{\odot}$ is the stellar mass, $r=R/R_{Sch}$ is the radius from the
star, and $R_{Sch}=2.95\times 10^5m~cm$ is the Schwartzschild radius
(e.g. Yi \& Narayan 1995, Rees et al. 1982).
The energetic protons could transfer their energies to electrons on the
electron-ion Coulomb exchange time scale
$t_{ie}\sim 9\times 10^{-5} \theta_e^{3/2}\alpha m {\dot m}^{-1} r^{3/2}~s$
where ${\dot m}={\dot M}/{\dot M}_{Edd}$ is the dimensionless accretion rate,
${\dot M}_{Edd}=1.39\times 10^{18}m~g/s$ is the Eddington accretion rate,
$\theta_e=kT_e/m_e c^2\simle $ a few is the dimensionless electron 
temperature, and $\alpha\sim 0.1$ is the dimensionless viscosity parameter
(e.g. Frank et al. 1992). 
If the magnetic field has the equipartition strength,
$t_{sync}/t_{acc}\sim 3\times 10^3\gamma_2^{-1}\alpha^2{\dot m}^{-1}r^2$ and 
$t_{sync}/t_{ie}\sim 9\times 10^2\theta_e^{-3/2}r$, which indicates that only
a very small fraction of the proton energy could be radiated by the
proton synchrotron emission as expected. However, if there exists an
external strong magnetic field such as that around a pulsar, the magnetic field
strength is $B_8\sim 4\times 10^5 \mu_{30} m^{-3} r^{-3}$ where
$\mu_{30}=\mu/10^{30}~Gcm^3$ is the magnetic moment of the star
(Frank et al. 1992).
In this case, $t_{sync}<t_{acc}$ occurs at $r\simle 20\gamma_2^{2/11}
\mu_{30}^{4/11}\alpha_{-1}^{-2/11}m^{-10/11}$ and $t_{sync}<t_{ie}$
occurs when 
$r\simle 50\gamma_2^{2/9}\theta_e^{1/3}\mu_{30}^{4/9}\alpha_{-1}^{2/9}
m^{-10/9}{\dot m}_{-2}^{-2/9}$ where $\alpha_{-1}=\alpha/0.1$. 
Therefore, in the region close to the neutron star, proton synchrotron
cooling could be a significant channel for proton cooling.
We suggest an observational signature based on the possible proton
synchrotron cooling near the strongly magnetized stars.
The optical emission near $5500{\AA}$ from 4U 1626-67 (Chakrabarty 1998)
appears interestingly close to the predicted synchrotron emission.

\section{High Energy Protons and Proton Synchrotron}

We assume that the hot accretion flow contains an equipartition strength
magnetic field (i.e. the gas pressure is equal to the magnetic pressure).
The relevant physical quantities are the equipartition magnetic field
$B\approx 8\times 10^8 \alpha^{-1/2} m^{-1/2} {\dot m}^{1/2} r^{-5/4}~G$
(e.g. Yi \& Narayan 1997), the ion temperature 
$T_i\approx 1\times 10^{12} r^{-1}~K$, and the proton number density
$n_p\approx 6\times 10^{19} \alpha^{-1} m^{-1} {\dot m} r^{-3/2} cm^{-3}$.
Although $T_i$ and $n_p$ quite plausibly represent the mean energy and
density of the protons, they do not constrain the possible non-thermal
relativistic proton population which could coexist with the non-relativistic
thermal protons.

The energy spectrum of the protons in the hot accretion flows is not well
understood. Since there is not a clear thermalization process working
for protons, if they are energized in non-thermal processes, their
energy distribution could remain non-thermal throughout accretion
(Narayan et al. 1998b). 
Non-thermal, power-law energy distributions for relativistic protons
have been recently motivated by possible gamma-ray emission and
low frequency radio emission signatures from Sgr A$^*$ 
(Mahadevan et al. 1997, Mahadevan 1998).
We take the number density of protons $n_p=\int n(\gamma,\theta_p)d\gamma$
where $\theta_p=kT_i/m_p c^2$. If protons have a fraction $f$ in the
non-thermal power-law tail while $1-f$ in the thermal, non-relativistic
($\gamma\sim 1$) protons, the fraction $f=3(s-2)\theta_p/2$ if the
mean energy of protons is $kT_i$ and the power-law slope is given by
$\propto \gamma^{-s}$. We have assumed that the thermal, non-relativistic
protons with $\gamma\sim 1$ exist as a separate proton population.
Since $\theta_p\sim 0.1r^{-1}$, $f\sim 0.2(s-2)r^{-1}$.
Of the total power-law protons, protons with $\gamma\simge 10$ are
energetic enough for synchrotron emission.
For $s=2.5$ (e.g. Mahadevan et al. 1997), at $r\sim 1$, $f\sim 0.1$ and
the fraction of protons with $\gamma\ge 10$ is $\sim 3\times 10^{-2}$. 
At $r\sim 10^2$, $f\sim 10^{-3}$. Therefore, the fraction of the relativistic 
protons with $\gamma>10$, $\epsilon$, is likely to be in the range 
$\sim 3\times 10^{-5}- \sim 3\times 10^{-4}$. Therefore, 
$\epsilon_{-4}=\epsilon/10^{-4}\sim 1$ could well represent the fraction 
of the relativistic non-thermal protons relevant for synchrotron emission
(cf. Mahadevan 1998).

The single particle synchrotron energy loss rate is (Lang 1980)
$\left|dE/dt\right|=3\times 10^{-2}\gamma_2^2B_8^2~erg/s$.
Using the effective number density of the relativistic protons 
$\epsilon n_p$, the total synchrotron energy loss rate in the entire
hot accretion flow is estimated as
\beq
L_{sync}\sim \int dR 4\pi R \left|dE/dt\right|\epsilon n_p
\sim 4\times 10^{33} \epsilon_{-4}\gamma_2^2\alpha^{-2} m {\dot m}^2
\eeq
where we have assumed that the magnetic field is the internal equipartition 
field. (A more detailed luminosity estimate involving integration over $\gamma$
follows below.) Since the hot accretion flows are likely to exist up to 
${\dot m}\sim 0.3\alpha^2$ (Narayan \& Yi 1995), the maximum synchrotron power 
is $L_{sync,max}\sim 3\times 10^{30} \epsilon_{-4}\gamma_2^2\alpha_{-1}^2 m$, 
which is a small fraction of the total viscously dissipated energy.

On the other hand, if the magnetic field provided by a dipole type field
of the neutron star $B_8=4\times 10^5 \mu_{30} m^{-3} r^{-3}$ (Frank et al.
1992),
$L_{sync}\sim 1\times 10^{39} \epsilon_{-4}\gamma_2^2\mu_{30}^2\alpha^{-1} 
m^{-4}{\dot m}~erg/s$. For the maximum accretion rate 
${\dot m}\sim 0.3\alpha^2$ for the two-temperature accretion flows 
(Yi \& Narayan 1995),
$L_{sync,max}\sim 3\times 10^{37} \epsilon_{-4}\gamma_2^2\mu_{30}^2\alpha_{-1}
m^{-4}~erg/s$. The external stellar field would be more important for
synchrotron cooling inside
the radius $r_s\sim 4\times 10^2 \mu_{30}^{4/7}\alpha_{-1}^{2/7} 
m^{-10/7}{\dot m}_{-2}^{-2/7}$ which is compared with the magnetospheric 
radius $r_o\sim 1\times 10^3\mu_{30}^{4/7}m^{4/7}{\dot m}_{-2}^{-2/7}$
(Frank et al. 1992, Yi et al. 1997, Yi \& Wheeler 1998). 
Therefore, emission inside the magnetospheric
radius could be dominated by the stellar field. If the accretion flows
are cooled rapidly inside the magnetospheric radius, then the
synchrotron luminosity could be entirely due to the accretion flow
present at $r>r_o$, which is estimated to be
$L_{sync}\sim 2\times 10^{30} \epsilon_{-4}\gamma_2^2\mu_{30}^2\alpha_{-1}^{-1}
m^{-4}{\dot m}_{-2}.$
This is comparable to the synchrotron luminosity due to
the internal equipartition field
$L_{sync}\sim 4\times 10^{31}\epsilon_{-4}\gamma^2\alpha_{-1}^{-2}
m{\dot m}_{-2}^2$.
Since the accretion flow inside the magnetospheric radius is adiabatically 
heated much like a spherical accretion flow, we assume that the accretion
flow contains a fraction $\epsilon\sim 10^{-4}$ of energetic, relativistic
protons until it hits the surface of the neutron star.

The synchrotron power at frequency $\nu$ for a single proton with the 
Lorentz factor $\gamma$ is (e.g. Lang 1980)
\beq
P(\nu,\gamma)={3^{1/2} e^3 B\sin\phi\over m_p c^2}{\nu\over \nu_c}
\int_{\nu\over \nu_c}^{\infty}dx K_{5/3}(x)
\eeq
where a major fraction of the synchrotron power is produced at the
characteristic frequency $\nu=\nu_c\nu_{cyc}\gamma^2=3eB\gamma^2/4\pi m_p c$
and $\sin\phi$ accounts for the pitch angle between magnetic field and
proton velocity. The emission coefficient is
\beq
j_{\nu}={1\over 4\pi}\int d\gamma N(\gamma)P(\nu,\gamma)
\eeq
and the absorption coefficient is
\beq
\alpha_\nu=-{1\over 8\pi m_p \nu^2}\int d\gamma P(\nu,\gamma){d\over d\gamma}
\left(N(\gamma)\over \gamma^2\right)
\eeq
where $N(\gamma)$ is the distribution of protons per unit volume with
the Lorentz factor $\gamma$. In our simple model for the relativistic
protons, $\int d\gamma N(\gamma) =\epsilon n_p$.

Taking a power-law slope $s=2.5$ and assuming a constant pitch angle
$\sin\phi=1/2$, we get
\beq
\alpha_{\nu}\approx 6.1\times 10^5 \epsilon n_p B^{9/4} \nu^{-13/4}~cm^{-1}
\eeq
or for a characteristic absorption length scale which is comparable
to the scale height of the hot accretion flow $H\sim R$
(e.g. Narayan \& Yi 1995), the self-absorption depth
$\tau_{\nu}\sim \alpha_{\nu} H\sim 6.1\times 10^5 \epsilon n_p R B^{9/4}
\nu^{-13/4}$.
Using the hot accretion flow solution,
$\tau_{\nu}\sim 2.2\times 10^{-3}\epsilon_{-4}\alpha_{-1}^{-1}{\dot m} 
r^{-1/2}B_8^{9/4} \nu_{15}^{-13/4}$
where $\epsilon_{-4}=\epsilon/10^{-4}$ and $\nu_{15}=\nu/10^{15} Hz$.
The source function for the self-absorbed part of the synchrotron emission
spectrum is
\beq
S_{\nu}=j_{\nu}/\alpha_{\nu}\sim 4\times 10^{-30} B^{-1/2}\nu^{5/2}~
erg/s/cm^2/Hz.
\eeq

When the magnetic field is the internal equipartition field,
\beq
\tau_{\nu}\sim 9\times 10^{-8}\epsilon_{-4}\alpha_{-1}^{-17/8}m^{-9/8}
{\dot m}_{-2}^{17/8} r_1^{-53/16} \nu_{15}^{-13/4}
\eeq
where $r_1=r/10$. The self-absorption occurs when $\tau_{\nu}=1$ or at
$\nu_{abs}\sim 7\times 10^{12} \epsilon_{-4}^{4/13}\alpha_{-1}^{-17/26}m^{-9/26}
{\dot m}_{-2}^{17/26} r_1^{-53/52} Hz$.
The synchrotron luminosity $L_{sync}=\nu L_{\nu}$ is estimated as
\beq
L_{sync}\sim 4\times 10^{26} \epsilon_{-4}^{0.79}\alpha_{-1}^{-1.43}m^{1.61}
{\dot m}_{-2}^{1.43}\nu_{13}^{0.92}~erg/s
\eeq
which could extend to
$\nu=\nu_{max}\sim 7\times 10^{13} \epsilon_{-4}^{0.31}\alpha_{-1}^{-0.65}
m^{-0.35}{\dot m}_{-2}^{0.65}$ Hz
where exponents have been rounded off for convenience. Therefore, such
a proton synchrotron luminosity is obviously several orders of magnitude 
lower than the luminosity due to electron cooling for all reasonable 
parameters (Yi \& Narayan 1995).

However, in strongly magnetized neutron star systems,
the magnetic field is the stellar field of the dipole type, 
$B_8=4\times 10^5 \mu_{30} m^{-3} r^{-3}$ (e.g. Frank et al. 1992)
and the synchrotron emission could be much stronger and the emission 
can extend to much higher frequency.
That is, the absorption depth 
$\tau_{\nu}\sim \epsilon_{-4}\mu_{30}^{9/13}\alpha_{-1}^{-1}m^{-29/4}
{\dot m}_{-2}r_1^{-8}\nu_{15}^{-13/4}$
and the self-absorption occurs at
\beq
\nu_{abs}\sim 1\times 10^{15} \epsilon_{-4}^{4/13}\mu_{30}^{9/13}
\alpha_{-1}^{-4/13} m^{-29/13}{\dot m}_{-2}^{4/13}r_1^{-32/13}~Hz.
\eeq
The luminosity is estimated as
$L_{sync}\sim 4\times 10^{33} \epsilon_{-4}^{0.44}\mu_{30}^{0.55}
\alpha_{-1}^{0.44}m^{0.33}{\dot m}_{-2}^{0.44}\nu_{15}^{2.08}~erg/s$.
The emission could extend up to
$\nu=\nu_{max}\sim 2\times 10^{16}\epsilon_{-4}^{0.31}\mu_{30}^{0.69}
\alpha_{-1}^{-0.31}m^{-2.23}{\dot m}_{-2}^{0.31}$ Hz
where the exponents have again been rounded off for convenience. 
The expected luminosity is significant enough for possible detection
(see below).

\section{Possible Detections}
We expect roughly $L_{sync}=\nu L_{\nu} \propto \nu$ for black hole systems
where the magnetic fields are internal equipartition type fields and 
$L_{sync}=\nu L_{\nu} \propto \nu^2$ for neutron star systems with strong
magnetic fields. Since the hot accretion flows are most likely during
low luminosity states, the above estimate for the internal field case
(for black holes) suggests that in black hole systems, detection of
the proton synchrotron emission is unlikely. For instance, during
quiescence of A0620-00 ($M\sim 6M_{\odot}$), 
${\dot m}\sim 2\times 10^{-4}$ for $\alpha\sim 0.3$ 
(Yi \& Narayan 1997 and references therein). Then, using the
above results we immediately get
$\nu L_{\nu}\sim 1\times 10^{25} \epsilon_{-4}^{0.79}\alpha_{-1}^{-1.43}~erg/s$
at $\nu=10^{13} Hz$ and
$\nu L_{\nu}\sim 1\times 10^{26} \epsilon_{-4}^{0.79}\alpha_{-1}^{-1.43}~erg/s$
at $\nu=10^{14} Hz$.
The proton synchrotron emission is too weak for detection. Even for accretion
rates as high as ${\dot m}\sim 0.1$, the synchrotron emission is unlikely to
be detected. In most cases, the electron synchrotron emission is expected
to be much stronger and it could have a significant contribution to the
observed emission spectra (Narayan \& Yi 1995).

The existence of the hot, two-temperature accretion flows in neutron star 
systems is hard to prove because the emission from the surface of the neutron
star dominates the emission spectrum (e.g. Narayan \& Yi 1995, Yi et al. 1996). 
The luminosity from the stellar surface
$L_x\sim GM{\dot M}/R_{NS}\sim 3\times 10^{36} {\dot m}_{-2}~erg/s$ is
most likely to occur in the X-ray range (Frank et al. 1992). 
Therefore, the luminosity from the neutron star systems is not expected 
to reflect the low radiative efficiency
which is often attributed to the hot, two-temperature accretion flows
(Narayan \& Yi 1995, Yi et al. 1996, Narayan et al. 1998b).
Using the expression for $L_x$, we get
\beq
L_{sync}\sim 2\times 10^{33} \epsilon_{-4}^{0.44}\mu_{30}^{0.55}
\alpha_{-1}^{0.44}m^{0.33}L_{x,36}^{0.44}\nu_{15}^{2.08}
\eeq
where $L_{x,36}=L_x/10^{36} erg/s$.
However, there are a few pulsar systems which show some indication of the hot 
accretion flows. That is, some systems have shown puzzling torque reversals (Yi et al. 1997). The spin-down episodes could be due to the transition of
accretion flows from cool, thin accretion disk to hot accretion flows
(Yi \& Wheeler 1998).
In the neutron star systems, the electron temperature becomes much lower than
that of the black hole systems due to intense cooling of electrons by
soft photons emitted at the surface of the neutron star where the
accretion flow lands. Since the electrons and ions are not strongly
coupled (only weakly through Coulomb coupling), the ion temperature
remains nearly unaffected. As a result, the electron temperature becomes much
lower than $\sim 10^9K$ and the electron synchrotron emission is
effectively quenched (Narayan \& Yi 1995). As long as the protons remain hot, 
the proton synchrotron emission remains unaffected.

Assuming that $M=1.4M_{\odot}$ and $R_{NS}=10 km$, we consider the pulsar
systems which showed abrupt torque reversals and consider their spin-down
episodes as due to the hot, two-temperature accretion flows.
First, 4U 1626-67's torque reversal event could be accounted for by
${\dot m}\sim 2\times 10^{-2}$ and $\mu_{30}\sim 2$ 
(Yi et al. 1997), which lead to
$L_{sync}\sim 9\times 10^{33} \epsilon_{-4}^{0.44}\alpha_{-1}^{0.44}
~erg/s$ at $\nu=10^{15} Hz$ and
$L_{sync}\sim 1\times 10^{36} \epsilon_{-4}^{0.44}\alpha_{-1}^{0.44}
~erg/s$ at $\nu=10^{16} Hz$. The soft X-ray luminosity is expected
at $L_x\sim 6\times 10^{36} erg/s$.
Similarly, GX 1+4's parameters are estimated as 
${\dot m}\sim 5\times 10^{-2}$ and $\mu_{30}\sim 50$ (Yi et al. 1997, Yi \& 
Wheeler 1998), which gives
$L_{sync}\sim 8\times 10^{34} \epsilon_{-4}^{0.44}\alpha_{-1}^{0.44}
~erg/s$ at $\nu=10^{15} Hz$ and
$L_{sync}\sim 9\times 10^{36} \epsilon_{-4}^{0.44}\alpha_{-1}^{0.44}
~erg/s$ at $\nu=10^{16} Hz$. 
The X-ray luminosity $L_x\sim 2\times 10^{37}~erg/s$
Finally, OAO 1657-415 (Yi \& Wheeler 1998) 
has also shown an abrupt torque reversal which
suggests ${\dot m}\sim 5\times 10^{-2}$ and $\mu_{30}\sim 20$ or
$L_{sync}\sim 4\times 10^{34} \epsilon_{-4}^{0.44}\alpha_{-1}^{0.44}
~erg/s$ at $\nu=10^{15} Hz$ and
$L_{sync}\sim 5\times 10^{36}\epsilon_{-4}^{0.44}\alpha_{-1}^{0.44}
~erg/s$ at $\nu=10^{16} Hz$. $L_x$ is expected to be similar to that
of GX 1+4.

It is particularly interesting that 4U 1626-67 has been detected in the
optical with a luminosity $\sim 5\times 10^{33}~erg/s$ at $5500{\AA}$
(for an assumed distance of 8.5 kpc) 
and $\nu L_{\nu}\propto \nu^2$ (Chakrabarty 1998), 
which is similar to our prediction, if the relativistic protons indeed 
have a power-law energy spectrum with index $s=2.5$. 
Given the fact that $s=2.5$ is amply motivated by the Galactic Center 
source Sgr A$^*$, it is quite interesting that a similar population of
relativistic protons may account for the optical emission in 4U 1626-67.
For the accretion parameters discussed above, if the observed optical emission
is the proton synchrotron emission, we immediately get an estimate on the
fraction of the relativistic protons 
$\epsilon\sim 5\times 10^{-4} \alpha_{-1}^{-1}$. This fraction is not
far from that assumed for the Galactic Center source Sgr A$^*$.
Chakrabarty (1998) suggests that the observed optical emission 
may be accounted for by the X-ray irradiated accretion flow. While this
possibility cannot be ruled out, it is difficult for a compact binary
such as 4U 1626-67 has a large outer radius required for the irradiation
to be effective in the accretion flow (Chakrabarty 1998). It is also
unclear whether the radiation from the magnetic poles can effectively
heat up the outer accretion flow (e.g. Yi \& Vishniac 1998).
In the case of GX 1+4, the pulsar is fed through wind accretion from a
giant secondary (e.g. Chakrabarty et al. 1998). 
It is unclear whether the torque reversing mechanism is
similar to that of 4U 1626-67 (Yi et al. 1997).

If the accretion flow at large radii is in the form of a thin disk,
the optical emission from the thin disk could occur at $\nu\sim 10^{15}
Hz$ with the quasi-blackbody spectra  at the temperature $\sim$ a few $10^4
{\dot m}_{-2} (r_o/10^3)^{-4/3}~K$ (Frank et al. 1992). 
The proton synchrotron luminosity
at $\nu\sim 10^{15} Hz$ exceeds the quasi-blackbody emission from the
thin disk if ${\dot m}\simle 2\times 10^{-2}(r_o/10^3)^{1.8}$. Therefore,
we conclude that the proton synchrotron emission should be seen if
the accretion flow is a two-temperature, hot flow with a characteristic
spectral index $\sim 1$ (i.e. $I_{\nu}\sim \nu$).

\section{Discussions}

We have shown that in strongly magnetized neutron star systems, the
existence of energetic protons in the hot accretion flows could be confirmed
by detection of synchrotron emission from relativistic protons. Although
the required number density of nonthermal protons remains highly uncertain,
the detection of such a radiation signature is possible if the relativistic 
protons are similar to those recently discussed in the context of the
gamma-ray emission from the Galactic center source Sgr A$^*$
(Mahadevan et al. 1997, Mahadevan 1998). The proton energy distribution
index $s=2.5$ makes a particularly interesting case in 4U 1626-67. A similar
index is motivated by the Sgr A$^*$'s proton signatures (Mahadevan et al.
1997, Mahadevan 1998).

The existence of the two-temperature plasma around accreting black holes and
neutron stars in their low luminosity states has been shown very plausible.
However, due to the lack of any direct test of such a possibility for 
relativistic protons, 
the spectral evidence has been the basis of the recent discussions
on the low efficiency, two-temperature flows. Therefore, non-detection of
the proton synchrotron signature would imply either that the hot protons
lack a relativistic component or that the neutron star systems do not have
the hot accretion flows. If the former is the case, the recently suggested
gamma-ray signature in Sgr $A^*$ could be questioned (Mahadevan et al.
1997). 

In neutron star systems, electrons are cooled much more efficiently than
in black hole systems while protons remain nearly virialized
(Narayan \& Yi 1995). Since ions
are not likely to be thermalized once they are produced by some nonthermal
acceleration processes, the relativistic protons could be highly nonthermal.
These protons could lose their energy significantly via proton synchrotron
emission. In black hole systems, due to lack of any strong fields, the
synchrotron emission is much weaker. Therefore, the two-temperature accretion 
flows and energetic protons could be "directly" detected more easily in neutron
star systems.

So far, there has not been a convincing evidence for the hot accretion
flows in the neutron star systems although the abrupt torque reversal
events have been attributed to the accretion flow transition between
the cool, geometrically thin accretion disk and the hot, geometrically
thick accretion flow. If the reversal is indeed due to the accretion
flow transition, the proton synchrotron emission would be seen only
during spin-down as the hot accretion flow exist only during spin-down.
Interestingly, it has been noted that the torque reversal seen in GX 1+4
is much more gradual and different from the more puzzling 4U 1626-67 event
(Yi et al. 1997). The detected optical emission in 4U 1626-67 shows the
luminosity and the spectral slope interestingly close to the proton
synchrotron emission. If the proton synchrotron is indeed responsible
for the optical emission, a strong polarization signal is expected.
If the GX 1+4 event is due to some mechanisms other than the accretion
flow transition, then GX 1+4's spin-down phase should lack the proton 
synchrotron emission signature. The predicted correlation between 
$L_{sync}$ and $L_x$ could provide an additional test for the 
two-temperature accretion flow.

\acknowledgments
IY thanks R. Mahadevan for related conversations and R. Narayan for informing of
a recent work on nuclear spallation. JY acknowledges a partial support from a
Ministry of Education research fund BSRI 97-2427, 
a MOST project 97-N6-01-01-A-9, and a KOSEF project 961-0210-061-2.
IY acknowledges a partial support from KRF grant 1998-001-D00365.

\vfill\eject
\clearpage


\begin{references}
\reference{A} Chakrabarty, D. 1998, ApJ, 492, 342
\reference{A} Chakrabarty, D., van Kerkwijk, M. H., \& Larkin, J. E. 1998,
ApJ, 497, L39
\reference{A} Frank, J., King, A. R., and Raine, D. 1992, Accretion Power 
in Astrophysics, Cambridge: Cambridge University Press
\reference{A} Lang, K. R. 1980, Astrophysical Formulae, Berlin: Springer-Verlag
\reference{A} Mahadevan, R., Narayan, R., \& Krolik, J. H. 1997, ApJ,
486, 268
\reference{A} Mahadevan, R. 1998, Nature, 394, 651
\reference{A} Manmoto, T., Mineshige, S., \& Kusunose, M. 1997, ApJ, 489, 791
\reference{A} Narayan, R., Mahadevan, R., Grindlay, J. E., Popham, R. G., \& 
gammie, C. 1998a, ApJ, 492, 554
\reference{A} Narayan, R., Mahadevan, R., \& Quataert, E. 1998b, preprint
(astro-ph/9803141)
\reference{A} Narayan, R. \& Yi, I. 1995, ApJ, 444, 231
\reference{A} Rees, M. J., Begelman, M. C., Blandford, R. D., Phinney, E. S.
1982, Nature, 295, 17
\reference{A} Rybicki, G. B. \& Lightman, A. P. 1980, 
Radiative Processes in Astrophysics, New York: John Wiley \& Sons
\reference{A} Yi, I. \& Boughn, S. P. 1998, ApJ, 499, 198
\reference{A} Yi, I. \& Narayan, R. 1997, ApJ, 486, 363
\reference{A} Yi, I., Narayan, R., Barret, D., \& McClintock, J. E. 1996,
A\&AS, 120C, 187
\reference{A} Yi, I. \& Vishniac, E. T. 1999, ApJ, in press
\reference{A} Yi, I. \& Wheeler, J. C. 1998, ApJ, 498, 802
\reference{A} Yi, I., Wheeler, J. C., \& Vishniac, E. T. 1997, ApJ, 481, L51
(Erratum ApJL, 491, L93)
\end{references}
\end{document}